\documentclass[aps,prb,twocolumn,superscriptaddress]{revtex4-1}
\usepackage{graphicx}
\usepackage{amssymb}
\usepackage{amsmath}
\usepackage{amsfonts}
\usepackage{braket}
\usepackage{color}

\begin{document}

\title{Dynamical susceptibility of Skyrmion crystal} 
\author{V.~E. Timofeev}
\email{vetimofeev@etu.ru}
\affiliation{NRC ``Kurchatov Institute", Petersburg Nuclear Physics Institute, Gatchina
188300, Russia}
\affiliation{St.\ Petersburg State University, 7/9 Universitetskaya nab., 199034
St.~Petersburg, Russia} 
\affiliation{St.\ Petersburg Electrotechnical University ``LETI'', 197376 St. Petersburg, Russia}
\author{D.~N. Aristov}
\affiliation{NRC ``Kurchatov Institute", Petersburg Nuclear Physics Institute, Gatchina
188300, Russia}
\affiliation{St.\ Petersburg State University, 7/9 Universitetskaya nab., 199034
St.~Petersburg, Russia} 
 
\begin{abstract}
Using stereographic projection approach we develop a theory for calculation of dynamical susceptibility tensor of Skyrmion crystals (SkX), formed in thin ferromagnetic films with Dzyaloshinskii-Moriya interaction and in the external magnetic field. Staying whenever possible within analytical framework, we employ the model anzats for static SkX configuration and discuss small fluctuations around it. The obtained formulas are numerically analyzed in the important case of uniform susceptibility, accessible in magnetic resonance (MR) experiments. We show that, in addition to three characteristic MR frequencies discussed earlier both theoretically and experimentally, one should also expect several resonances of smaller amplitude at somewhat higher frequencies.   
\end{abstract}

\maketitle

{\bf Introduction}.
Magnetic skyrmions are topologically protected particle-like configurations of local magnetization,  appearing particularly in non-centrosymmetric magnets\cite{bogdanov1989thermodynamically} with Dzyaloshinskii-Moriya interaction (DMI). Skyrmions are studied as perspective building blocks for novel computer memory devices\cite{Vakili_2021}, programmable logic devices\cite{Yan2021}, or even artificial neural network devices \cite{li2021magnetic}. It is well known that skyrmions are arranged into regular lattices \cite{muhlbauer2009skyrmion, Yu2010b}.  Skyrmion lattices, named also skyrmion crystals (SkX), attract the attention of researchers because of their applications to magnonics \cite{garst2017collective}.

One skyrmion can largely be considered as a small size magnetic bubble, whose motion can be described by Thiele equation\cite{thiele1973}. But even a single skyrmion is a complex structure, which has  its own dynamics that cannot be described in terms of skyrmion's displacement only. There are also  deformations of the skyrmion's form, such as dilatation, elliptical distortions, triangular distortions etc.\cite{schutte2014magnon,lin2014internal} 

It was shown that the energy band structure of SkX  should possess a Goldstone mode \cite{petrova2011},  associated with the displacement of skyrmions' centers. Besides this mode there are many other branches of different  symmetry \cite{mochizuki2012, garst2017collective,Diaz2020}, associated with, e.g.,  elliptical deformation, clockwise (CW) rotation, counterclockwise (CCW) rotation and breathing mode (Br) of skyrmions. 

One can observe and explore SkX excitations by several methods, among them inelastic neutron scattering\cite{Weber_2022}, optical inverse-Faraday effect \cite{ogawa2015ultrafast}, magnetic resonance (MR) technique. In the latter technique  it was shown   \cite{onose2012observation} that CW and CCW modes are observed when oscillating component of magnetic field is directed in the plane of SkX and the breathing mode is observed for the field perpendicular to the plane, in accordance with the earlier prediction by means of numerical simulations \cite{mochizuki2012}.

It was also demonstrated that other (octupole and sextupole) modes can manifest themselves in MR experiments in bulk SkX systems with strong cubic magnetocrystalline anisotropy, which hybridizes these modes  with Br and CCW excitations\cite{takagi2021,Aqeel2021}.

In this study we discuss the MR response of SkX formed in thin films with DMI and in presence of magnetic field at low temperatures. We find that  beyond the lowest-energy CW, CCW and Br modes, there are also higher-energy modes of the same symmetry. Therefore they  should also be visible in MR response experiment, although with a much smaller the magnitude of the corresponding signals. 


{\bf Model}. 
We consider a thin film of Heisenberg ferromagnet with Dzyaloshinskii-Moriya interaction an in uniform magnetic field, $B$, perpendicular to the film. The energy density is given by 
\begin{equation}
\mathcal{E} =   \frac{C}{2}  \partial_{\mu}S_{i}\partial_{\mu}S_{i} - 
D\epsilon_{\mu ij} S_{i}\partial_{\mu}S_{j}  - B  S_{3},
\label{classicalenergy}
\end{equation}
with $C$ and $D$ are exchange and DMI constants, respectively. This is perhaps the minimal model where the SkX exists (apart from centrosymmetric frustrated magnets, see \cite{Utesov2021, Utesov2022}) and we omit possible anisotropy terms in further discussion.  We  take the  low temperature limit, when the local magnetization  is saturated to its maximum value $\mathbf{S} = S \mathbf{n}$, and $|\mathbf{n}|=1$.  It is convenient to measure length in units of $l=C/D$, and energy density in units of $ CS^2 l^{-2} = S^2D^2/C$. Then the energy density \eqref{classicalenergy} becomes dependent only on the dimensionless field $b=BC /SD^2$. 
 It turns out that in a range $0.25 \alt b \alt 0.8$ the static configuration for $\mathbf{n}$ corresponds to SkX, extensively discussed in the literature.

In the stereographic projection method we represent the unit vector $\mathbf{n}$ along the local magnetization as 
\begin{equation}
n_1 + i n_2  = \frac{2f}{1 + f\bar{f}}\,,\quad 
n_3 = \frac{1 - f\bar{f}}{1 + f\bar{f}},
\label{eq:stereo}
\end{equation}
where $f=f(z,\bar{z})$ is a function of a complex variable $z=x+iy$ and a conjugate one $\bar{z}=x - iy$;  $x,y$ are spatial coordinates.  The expression for $\mathcal{E}$ acquires highly nonlinear form in terms of $f, \bar f$, presented elsewhere \cite{Timofeev_2022}.
 
We consider the dynamics of the local magnetization in  Lagrangian formalism, $\mathcal{L} = \mathcal{T} - \mathcal{E}$, with the kinetic term given by 
\begin{equation}
\mathcal{T}=
\frac { S}{ \gamma_0}(1-\cos{\theta})\dot{\varphi}\,,
\label{eq:kinetic}
\end{equation}
here  $\varphi$ and $\theta$ define the  magnetization direction $\mathbf{n}=(\cos\varphi\sin\theta,\sin\varphi\sin\theta,\cos\theta)$. This leads to the Landau-Lifshitz  equation, 
$\dot{\mathbf{S}}=-\gamma_0 \, \mathbf{S}\times\mathbf{H},$
with  $\gamma_0$ is a gyromagnetic ratio,   and $\mathbf{H}=\delta {E}/\delta\mathbf{S}$ is an effective magnetic field.

Absorbing the factor $S/\gamma_0$ into the time redefinition, the kinetic term may be written in terms of the complex function $f$ as 
\begin{equation}
\mathcal{T}[f]=  \frac i2 \frac{\bar f \partial_t f - f \partial _t \bar f}
{1+f \bar f}.
\label{kinLagrangian}
\end{equation}


We study the dynamics of local magnetization by considering small fluctuations of  the stereographic function, $f(\mathbf{r},t)$, around the static field, $f_0(\mathbf{r})$, providing the minimum of total energy,  $\int d\mathbf{r}\,\mathcal{E}$, and write 
\begin{equation}
f(\mathbf{r},t) = f_0(\mathbf{r}) + \alpha(1 + f_0(\mathbf{r}) \bar{f}_0(\mathbf{r}))\psi(\mathbf{r},t)  \,,  
\end{equation}
with $\psi$ is time-dependent function and  $\alpha$ is a small parameter of the theory, clarified below. We then consider the expansion $\mathcal{L} = \mathcal{L}_0+ \alpha \mathcal{L}_1 + \alpha^2 \mathcal{L}_2 +\ldots$. 

{\bf Normal modes}.
For properly chosen function $f_0$  the first order terms  are absent, $\mathcal{L}_1 =0$,   and 
 the second order terms are  
\begin{equation}
\mathcal{L}_2 =\frac{1}{2} 
\begin{pmatrix}
  \bar{\psi},& \psi
\end{pmatrix}
\left(-i
\begin{pmatrix}
  \partial_t& 0\\
  0& -\partial_t
\end{pmatrix}
-\hat{\mathcal{H}} \right)
\begin{pmatrix}
  \psi\\
  \bar{\psi}
\end{pmatrix},
\label{Lagr2}
\end{equation}
with the Hamiltonian operator $\hat{\mathcal{H}}$ of the form 
\begin{equation}
\hat{\mathcal{H}}=
\begin{pmatrix}
  (-i\nabla + \mathbf{A})^2 + U&
   V\\
  V^*&
   (i\nabla + \mathbf{A})^2 + U
\end{pmatrix} \,,
\label{ham}
\end{equation}
with $\nabla =  \mathbf{e}_x \partial_x + \mathbf{e}_y \partial_y$. 
Here $U$, $V$ and $\mathbf{A} = \mathbf{e}_x A_x + \mathbf{e}_y A_y$ are rather cumbersome  functions of $f_0(\mathbf{r})$ and its gradients, see \cite{Timofeev_2022}. 


The Lagrangian \eqref{Lagr2}  results in 
the Euler-Lagrange equation 
\begin{equation}
-i \frac{d}{dt}
\begin{pmatrix}
  \psi\\
  \bar{\psi}
\end{pmatrix}
=\sigma_3\hat{\mathcal{H}}
\begin{pmatrix}
  \psi\\
  \bar{\psi}
\end{pmatrix},
\label{BdGeq}
\end{equation}
and the energies of normal modes, $ \epsilon_n$, are found from  
\begin{equation}
\Big( \epsilon_n \, \sigma_3 - \hat{\mathcal{H}} \Big) 
\begin{pmatrix}
  u_n\\
  v_n
\end{pmatrix}  
=  0 \,, 
\label{eq:shr}
\end{equation}
with $\sigma_3$ the third Pauli matrix. The solutions to Eq. \eqref{eq:shr} are discussed at length in \cite{Timofeev_2022}. 
They form the complete orthonormal basis, with $  \int d^2\mathbf{r} \,\left(   |u_n|^2 -  | v_n|^2\right) = 1$, 
which allows to expand every function $\psi$ (together with its conjugate, $\bar \psi$) as 
\begin{equation}
\begin{pmatrix}
  \psi\\
  \bar{\psi}
\end{pmatrix} =  \sum\limits_n
\begin{pmatrix}
 u_n,&  v_n^*\\
v_n,&  u_n^*
\end{pmatrix}
\begin{pmatrix}
  c_n   \\
   c_n^\dagger
\end{pmatrix}
\label{psiexpan}
\end{equation}
here $c_n$ and $c_n^\dagger$ are mutually complex-conjugated numbers, which become boson creation and annihilation operators upon the second quantization procedure. 
The Hamiltonian then takes the familiar form, $\sum _n \epsilon_n  (c_n^\dagger c_n+1/2)$. 


{\bf Magnetization expansion}. 
The   expansion in $\alpha$
 of local magnetization is given by:
\begin{equation}
S_i =S_i^{(0)} + 
\alpha S  ( \bar{F}_i  \psi +  F_i   \bar{\psi})
 + O(\alpha^2),
\label{Slinear}
\end{equation}
with $S_i^{(0)} = S n_i$  defined by   \eqref{eq:stereo} with $f=f_0$, and 
\begin{equation}
\mathbf{F}=\frac{1}{1 + f_0\bar{f}_0}
\begin{pmatrix}
  1-f_0^2\\
  i(1 + f_0^2)\\
  -2 f_0
\end{pmatrix}
\label{def:F}
\end{equation}
is a complex $\mathbf{r}$-dependent vector with the following property: three vectors $\mbox{Re }\mathbf{F}$, $\mbox{Im }\mathbf{F}$, and $\mathbf{n}$ form the orthonormal basis. 


Substituting (\ref{psiexpan}) into the (\ref{Slinear}) we write to the  lowest order  in $\alpha$
\begin{equation}
S_{i} = S_i^{(0)} + \alpha S \sum\limits_n\big((\bar{F}_iu_n + F_iv_n)c_n + H.c.\big)
\end{equation}
Using  canonical commutation relations, $[c_n, c_m^\dagger]=\delta _{nm}$, and the completeness of the basis $(u_n,v_n)$,  one can check that taking the value $\alpha=1/\sqrt{2S}$ we obtain the 
  relation  
\begin{equation}
[S_{j}(\mathbf{r}),S_{k}(\mathbf{r}')] =  i \epsilon_{jkl}S_l^{(0)}(\mathbf{r}) \,\delta(\mathbf{r}-\mathbf{r}') \,, 
\end{equation}
which is expected in the linear spin-wave theory. 



{\bf Susceptibility}. 
Our aim is to calculate the dynamic susceptibility tensor, $\chi_{ij}(\mathbf{k},\omega) =\int dt\,e^{i\omega t} \chi_{ij}(\mathbf{k},t)$, which is the Fourier transform of  the spin retarded Green's function 
\begin{equation}
\chi_{ij}(\mathbf{k},t) = -i \theta(t)\langle[S_i(\mathbf{k},t),S_{j}(-\mathbf{k},0)]\rangle
\end{equation}

Using the above formulas, we can write 
\begin{equation}
S_{i}(\mathbf{k},t) =   S_i^{(0)} (\mathbf{k}) +  \sqrt{\frac{S}2}\sum\limits_n (A^i_n(\mathbf{k})e^{-i\epsilon_{n } t}c_n + H.c.)\, .
\end{equation}
The amplitude of the spin wave with the energy $\epsilon_n$ and the wave-vector $\mathbf{k}$ is given by 
\begin{equation}
A_n^j(\mathbf{k}) = \int d\mathbf{r}\, e^{i\mathbf{k}\mathbf{r}}(\bar{F}_ju_n + F_jv_n) \,. 
\label{def:Anj}
\end{equation}
As a result we obtain the general expression
\begin{equation}
\chi_{ij}(\mathbf{k},\omega) = \frac{S}{2}\sum\limits_n\left(\frac{\bar{A}^i_n(\mathbf{k})A^j_n(-\mathbf{k})}{\omega+\epsilon_n +i\delta} - \frac{A^i_n(\mathbf{k})\bar{A}^j_n(-\mathbf{k})}{\omega-\epsilon_n +i\delta}\right)
\,. 
\end{equation}
We should note here that  index $n$ of the mode  $\epsilon_n$ assumes both the wave-vector $\mathbf{k}$ and the number of the magnon band, see \cite{Timofeev_2022}.   In the important case of $\mathbf{k}=0$, relevant to magnetic resonance experiments and discussed below, the above formula is somewhat simplified. 

{\em Uniform susceptibility}. 
We set  $\mathbf{k}=0$ and denote $A^j_n(0)\equiv A_{j,n}$. Generally we expect that the tensor $\chi_{ij}$ contains symmetric and antisymmetric part, and the only chosen direction in our problem, Eq. \eqref{classicalenergy},  is normal to the plane, $\hat h = (0,0,1)$. We can then write  : 
\begin{equation}
\begin{aligned}
\chi_{ij}(\omega) &= \chi_\| \hat h_i \hat h_j + \chi_\perp (\delta_{ij} - \hat h_i \hat h_j) +
\chi_{as } \epsilon_{ijk}\hat h_k \,,
\\
 \chi_\| &= 
  \sum\limits_n \frac{ \epsilon_n  S\, |A_{3,n} |^2 }{-(\omega +i\delta)^2+\epsilon_n^2}  \,, 
\\
\chi_\perp &= 
\sum\limits_n \frac{ \epsilon_n S\, | A_{1,n}|^2}{-(\omega +i\delta)^2+\epsilon_n^2}  \,, 
\\
\chi_{as }  &= 
 \sum\limits_n \frac{i\omega S \mbox{ Im}({A}_{1,n}\bar{A}_{2,n} ) }{-(\omega +i\delta)^2+\epsilon_n^2}  \,,
%
\end{aligned}
\label{chi-comp}
\end{equation}

\begin{figure}[t]
\center{\includegraphics[width=0.99\linewidth]{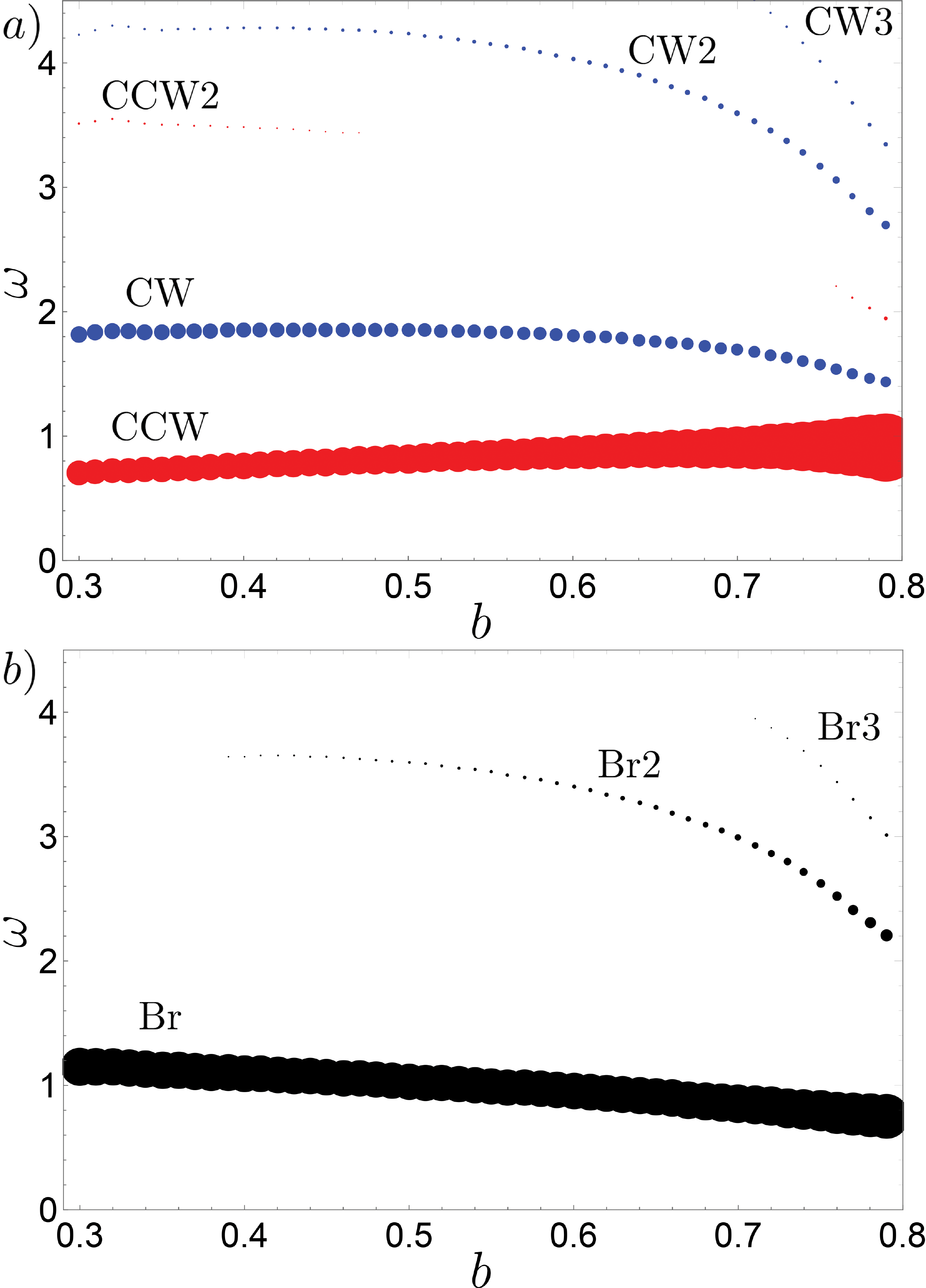}}
\caption{Imaginary part of a) transverse, $\chi_\perp$, and b) longitudinal, $\chi_\|$, components of a susceptibility \eqref{chi-comp} as a function of magnetic field $b$. The area of each circle is proportional to the weight, $|A_{j,n}|^2$, of each delta-function. Three well known lowest resonances: breathing (Br), clockwise (CW) and counterclockwise (CCW) are clearly visible, together with  higher energy modes of smaller intensity.}
\label{fig:sus}
\end{figure}

{\bf Susceptibility of SkX}. 
Using our  ansatz for static SkX configuration $f_0$, discussed in \cite{Timofeev2019,Timofeev_2022}, we calculate the magnon wave functions corresponding to $\mathbf{k}=0$ in \eqref{eq:shr}, for 36 lowest energies, $\epsilon_n>0$. The above formulas \eqref{def:F} and \eqref{def:Anj} are combined for calculation of susceptibility components, Eq. \eqref{chi-comp}.

The results of this calculation can be summarized as follows. 

\noindent
i. Only some of the normal modes of the spectrum are visible, thanks to selection rules, provided by the matrix elements $A_{j,n}$ in \eqref{chi-comp}. Specifically, the symmetry of the corresponding magnon wave function at the center of skyrmions, $\psi \sim z^m$, defines this visibility. The modes with magnetic quantum numbers $m=0, 2$ show up in $\chi_\perp$ and the modes with $m=1$ define $\chi_\|$. 
The lowest-energy modes with $m=0,1,2$ were dubbed  ``counterclockwise'' (CCW), ``breathing'' (Br) and  ``clockwise'' (CW), respectively, thanks to their dynamical pattern discussed elsewhere \cite{mochizuki2012,Timofeev_2022}. 

\noindent 
ii. In case of CCW and CW modes the antisymmetric part of susceptibility, $\chi_{as }$, is equal in absolute value to its diagonal part, $\chi_{\perp}$, with the residues $\mbox{ Im}({A}_{1,n}\bar{A}_{2,n} ) =-|A_{1,n}|^2 $ and $\mbox{ Im}({A}_{1,n}\bar{A}_{2,n} ) = |A_{1,n}|^2 $, respectively. 

\noindent
iii. In contrast to previous studies, we observe several modes with increasing energies for each $m$, which can in principle be observed experimentally. Generally, the increase of $\epsilon_n$ is accompanied  by the decrease of the weight of the corresponding resonance, i.e. the residue $|A_{j,n}|^2$ in \eqref{chi-comp}. We graphically represent the position of the MR lines and their weight in Fig.\ \ref{fig:sus}.

\noindent
iv. It is seen in this figure that the most intense lines correspond to lowest frequencies, which have a tendency to decrease with the external magnetic field, $b$. Labeling CW (CCW) in Fig.\ \ref{fig:sus}a is done according to sign of $\chi_{as }$, as explained above.

\begin{figure}[t]
\center{\includegraphics[width=0.99\linewidth]{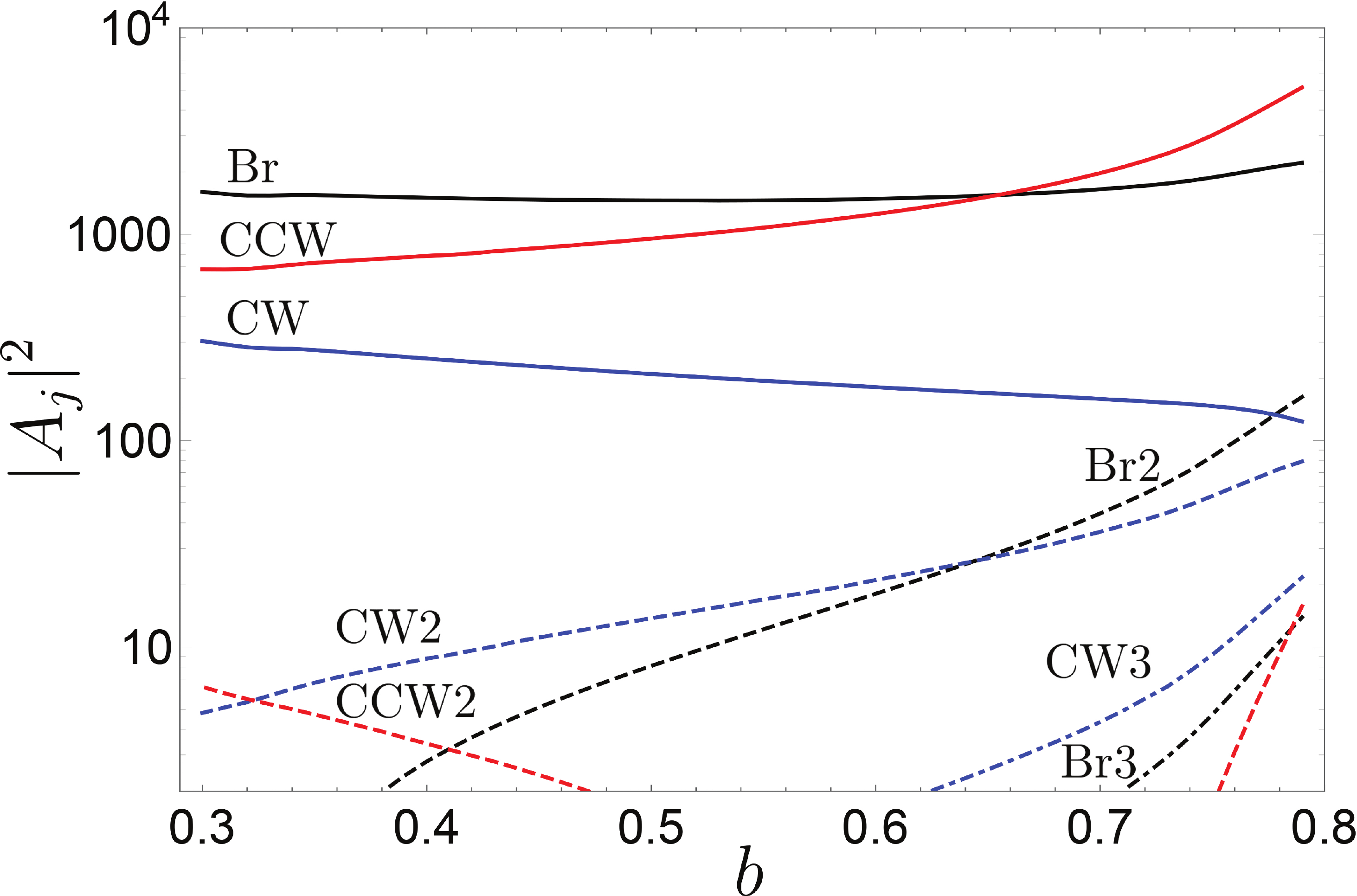}}
\caption{Spectral weight of eight resonances depicted in Fig.\ref{fig:sus} as a function of magnetic field $b$.}
\label{fig:A2}
\end{figure}

\noindent
v. The intensity of the lines, $|A_{j,n}|^2$, in \eqref{chi-comp} are plotted separately in Fig.\ \ref{fig:A2} as a function of applied field. We see that at higher fields, $b\simeq 0.75$, close to the melting point of SkX, the intensities of secondary CW2 and Br2 modes become comparable to the intensity of the main CW mode. This prediction would be interesting to check experimentally.  

Summarizing, we present the theory of the dynamical susceptibility of skyrmion crystal in the framework of stereographic projection approach. The obtained formulas are quite general and do not assume a specific type of skyrmion ordering. Applying our theory to hexagonal lattice of Bloch-type skyrmions we show the existence of several resonant frequencies, of which only three lowest ones were previously discussed in the literature. 
   
{\bf Acknowledgements}. 
The work was supported by the Russian Science Foundation, Grant No. 22-22-20034 and St.Petersburg Science Foundation, Grant No. 33/2022. 
The work of  V.T. was partially supported by the Foundation for the Advancement of Theoretical Physics BASIS. 

\bibliography{skyrmionbib}

\begin{thebibliography}{22}%
\makeatletter
\providecommand \@ifxundefined [1]{%
 \@ifx{#1\undefined}
}%
\providecommand \@ifnum [1]{%
 \ifnum #1\expandafter \@firstoftwo
 \else \expandafter \@secondoftwo
 \fi
}%
\providecommand \@ifx [1]{%
 \ifx #1\expandafter \@firstoftwo
 \else \expandafter \@secondoftwo
 \fi
}%
\providecommand \natexlab [1]{#1}%
\providecommand \enquote  [1]{``#1''}%
\providecommand \bibnamefont  [1]{#1}%
\providecommand \bibfnamefont [1]{#1}%
\providecommand \citenamefont [1]{#1}%
\providecommand \href@noop [0]{\@secondoftwo}%
\providecommand \href [0]{\begingroup \@sanitize@url \@href}%
\providecommand \@href[1]{\@@startlink{#1}\@@href}%
\providecommand \@@href[1]{\endgroup#1\@@endlink}%
\providecommand \@sanitize@url [0]{\catcode `\\12\catcode `\$12\catcode
  `\&12\catcode `\#12\catcode `\^12\catcode `\_12\catcode `\%12\relax}%
\providecommand \@@startlink[1]{}%
\providecommand \@@endlink[0]{}%
\providecommand \url  [0]{\begingroup\@sanitize@url \@url }%
\providecommand \@url [1]{\endgroup\@href {#1}{\urlprefix }}%
\providecommand \urlprefix  [0]{URL }%
\providecommand \Eprint [0]{\href }%
\providecommand \doibase [0]{http://dx.doi.org/}%
\providecommand \selectlanguage [0]{\@gobble}%
\providecommand \bibinfo  [0]{\@secondoftwo}%
\providecommand \bibfield  [0]{\@secondoftwo}%
\providecommand \translation [1]{[#1]}%
\providecommand \BibitemOpen [0]{}%
\providecommand \bibitemStop [0]{}%
\providecommand \bibitemNoStop [0]{.\EOS\space}%
\providecommand \EOS [0]{\spacefactor3000\relax}%
\providecommand \BibitemShut  [1]{\csname bibitem#1\endcsname}%
\let\auto@bib@innerbib\@empty
\bibitem [{\citenamefont {Bogdanov}\ and\ \citenamefont
  {Yablonskii}(1989)}]{bogdanov1989thermodynamically}%
  \BibitemOpen
  \bibfield  {author} {\bibinfo {author} {\bibfnamefont {A.~N.}\ \bibnamefont
  {Bogdanov}}\ and\ \bibinfo {author} {\bibfnamefont {D.}~\bibnamefont
  {Yablonskii}},\ }\href@noop {} {\bibfield  {journal} {\bibinfo  {journal}
  {Zh. Eksp. Teor. Fiz}\ }\textbf {\bibinfo {volume} {95}},\ \bibinfo {pages}
  {178} (\bibinfo {year} {1989})}\BibitemShut {NoStop}%
\bibitem [{\citenamefont {Vakili}\ \emph {et~al.}(2021)\citenamefont {Vakili},
  \citenamefont {Xu}, \citenamefont {Zhou}, \citenamefont {Sakib},
  \citenamefont {Morshed}, \citenamefont {Hartnett}, \citenamefont {Quessab},
  \citenamefont {Litzius}, \citenamefont {Ma}, \citenamefont {Ganguly},
  \citenamefont {Stan}, \citenamefont {Balachandran}, \citenamefont {Beach},
  \citenamefont {Poon}, \citenamefont {Kent},\ and\ \citenamefont
  {Ghosh}}]{Vakili_2021}%
  \BibitemOpen
  \bibfield  {author} {\bibinfo {author} {\bibfnamefont {H.}~\bibnamefont
  {Vakili}}, \bibinfo {author} {\bibfnamefont {J.-W.}\ \bibnamefont {Xu}},
  \bibinfo {author} {\bibfnamefont {W.}~\bibnamefont {Zhou}}, \bibinfo {author}
  {\bibfnamefont {M.~N.}\ \bibnamefont {Sakib}}, \bibinfo {author}
  {\bibfnamefont {M.~G.}\ \bibnamefont {Morshed}}, \bibinfo {author}
  {\bibfnamefont {T.}~\bibnamefont {Hartnett}}, \bibinfo {author}
  {\bibfnamefont {Y.}~\bibnamefont {Quessab}}, \bibinfo {author} {\bibfnamefont
  {K.}~\bibnamefont {Litzius}}, \bibinfo {author} {\bibfnamefont {C.~T.}\
  \bibnamefont {Ma}}, \bibinfo {author} {\bibfnamefont {S.}~\bibnamefont
  {Ganguly}}, \bibinfo {author} {\bibfnamefont {M.~R.}\ \bibnamefont {Stan}},
  \bibinfo {author} {\bibfnamefont {P.~V.}\ \bibnamefont {Balachandran}},
  \bibinfo {author} {\bibfnamefont {G.~S.~D.}\ \bibnamefont {Beach}}, \bibinfo
  {author} {\bibfnamefont {S.~J.}\ \bibnamefont {Poon}}, \bibinfo {author}
  {\bibfnamefont {A.~D.}\ \bibnamefont {Kent}}, \ and\ \bibinfo {author}
  {\bibfnamefont {A.~W.}\ \bibnamefont {Ghosh}},\ }\href {\doibase
  10.1063/5.0046950} {\bibfield  {journal} {\bibinfo  {journal} {Journal of
  Applied Physics}\ }\textbf {\bibinfo {volume} {130}},\ \bibinfo {pages}
  {070908} (\bibinfo {year} {2021})}\BibitemShut {NoStop}%
\bibitem [{\citenamefont {Yan}\ \emph {et~al.}(2021)\citenamefont {Yan},
  \citenamefont {Liu}, \citenamefont {Guang}, \citenamefont {Yue},
  \citenamefont {Feng}, \citenamefont {Lake}, \citenamefont {Yu},\ and\
  \citenamefont {Han}}]{Yan2021}%
  \BibitemOpen
  \bibfield  {author} {\bibinfo {author} {\bibfnamefont {Z.}~\bibnamefont
  {Yan}}, \bibinfo {author} {\bibfnamefont {Y.}~\bibnamefont {Liu}}, \bibinfo
  {author} {\bibfnamefont {Y.}~\bibnamefont {Guang}}, \bibinfo {author}
  {\bibfnamefont {K.}~\bibnamefont {Yue}}, \bibinfo {author} {\bibfnamefont
  {J.}~\bibnamefont {Feng}}, \bibinfo {author} {\bibfnamefont {R.}~\bibnamefont
  {Lake}}, \bibinfo {author} {\bibfnamefont {G.}~\bibnamefont {Yu}}, \ and\
  \bibinfo {author} {\bibfnamefont {X.}~\bibnamefont {Han}},\ }\href {\doibase
  10.1103/physrevapplied.15.064004} {\bibfield  {journal} {\bibinfo  {journal}
  {Physical Review Applied}\ }\textbf {\bibinfo {volume} {15}},\ \bibinfo
  {pages} {064004} (\bibinfo {year} {2021})}\BibitemShut {NoStop}%
\bibitem [{\citenamefont {Li}\ \emph {et~al.}(2021)\citenamefont {Li},
  \citenamefont {Kang}, \citenamefont {Zhang}, \citenamefont {Nie},
  \citenamefont {Zhou}, \citenamefont {Wang},\ and\ \citenamefont
  {Zhao}}]{li2021magnetic}%
  \BibitemOpen
  \bibfield  {author} {\bibinfo {author} {\bibfnamefont {S.}~\bibnamefont
  {Li}}, \bibinfo {author} {\bibfnamefont {W.}~\bibnamefont {Kang}}, \bibinfo
  {author} {\bibfnamefont {X.}~\bibnamefont {Zhang}}, \bibinfo {author}
  {\bibfnamefont {T.}~\bibnamefont {Nie}}, \bibinfo {author} {\bibfnamefont
  {Y.}~\bibnamefont {Zhou}}, \bibinfo {author} {\bibfnamefont {K.~L.}\
  \bibnamefont {Wang}}, \ and\ \bibinfo {author} {\bibfnamefont
  {W.}~\bibnamefont {Zhao}},\ }\href@noop {} {\bibfield  {journal} {\bibinfo
  {journal} {Materials Horizons}\ }\textbf {\bibinfo {volume} {8}},\ \bibinfo
  {pages} {854} (\bibinfo {year} {2021})}\BibitemShut {NoStop}%
\bibitem [{\citenamefont {M{\"u}hlbauer}\ \emph {et~al.}(2009)\citenamefont
  {M{\"u}hlbauer}, \citenamefont {Binz}, \citenamefont {Jonietz}, \citenamefont
  {Pfleiderer}, \citenamefont {Rosch}, \citenamefont {Neubauer}, \citenamefont
  {Georgii},\ and\ \citenamefont {B{\"o}ni}}]{muhlbauer2009skyrmion}%
  \BibitemOpen
  \bibfield  {author} {\bibinfo {author} {\bibfnamefont {S.}~\bibnamefont
  {M{\"u}hlbauer}}, \bibinfo {author} {\bibfnamefont {B.}~\bibnamefont {Binz}},
  \bibinfo {author} {\bibfnamefont {F.}~\bibnamefont {Jonietz}}, \bibinfo
  {author} {\bibfnamefont {C.}~\bibnamefont {Pfleiderer}}, \bibinfo {author}
  {\bibfnamefont {A.}~\bibnamefont {Rosch}}, \bibinfo {author} {\bibfnamefont
  {A.}~\bibnamefont {Neubauer}}, \bibinfo {author} {\bibfnamefont
  {R.}~\bibnamefont {Georgii}}, \ and\ \bibinfo {author} {\bibfnamefont
  {P.}~\bibnamefont {B{\"o}ni}},\ }\href {\doibase 10.1126/science.1166767}
  {\bibfield  {journal} {\bibinfo  {journal} {Science}\ }\textbf {\bibinfo
  {volume} {323}},\ \bibinfo {pages} {915} (\bibinfo {year}
  {2009})}\BibitemShut {NoStop}%
\bibitem [{\citenamefont {Yu}\ \emph {et~al.}(2010)\citenamefont {Yu},
  \citenamefont {Onose}, \citenamefont {Kanazawa}, \citenamefont {Park},
  \citenamefont {Han}, \citenamefont {Matsui}, \citenamefont {Nagaosa},\ and\
  \citenamefont {Tokura}}]{Yu2010b}%
  \BibitemOpen
  \bibfield  {author} {\bibinfo {author} {\bibfnamefont {X.~Z.}\ \bibnamefont
  {Yu}}, \bibinfo {author} {\bibfnamefont {Y.}~\bibnamefont {Onose}}, \bibinfo
  {author} {\bibfnamefont {N.}~\bibnamefont {Kanazawa}}, \bibinfo {author}
  {\bibfnamefont {J.~H.}\ \bibnamefont {Park}}, \bibinfo {author}
  {\bibfnamefont {J.~H.}\ \bibnamefont {Han}}, \bibinfo {author} {\bibfnamefont
  {Y.}~\bibnamefont {Matsui}}, \bibinfo {author} {\bibfnamefont
  {N.}~\bibnamefont {Nagaosa}}, \ and\ \bibinfo {author} {\bibfnamefont
  {Y.}~\bibnamefont {Tokura}},\ }\href {\doibase 10.1038/nature09124}
  {\bibfield  {journal} {\bibinfo  {journal} {Nature}\ }\textbf {\bibinfo
  {volume} {465}},\ \bibinfo {pages} {901} (\bibinfo {year}
  {2010})}\BibitemShut {NoStop}%
\bibitem [{\citenamefont {Garst}\ \emph {et~al.}(2017)\citenamefont {Garst},
  \citenamefont {Waizner},\ and\ \citenamefont
  {Grundler}}]{garst2017collective}%
  \BibitemOpen
  \bibfield  {author} {\bibinfo {author} {\bibfnamefont {M.}~\bibnamefont
  {Garst}}, \bibinfo {author} {\bibfnamefont {J.}~\bibnamefont {Waizner}}, \
  and\ \bibinfo {author} {\bibfnamefont {D.}~\bibnamefont {Grundler}},\ }\href
  {\doibase 10.1088/1361-6463/aa7573} {\bibfield  {journal} {\bibinfo
  {journal} {Journal of Physics D: Applied Physics}\ }\textbf {\bibinfo
  {volume} {50}},\ \bibinfo {pages} {293002} (\bibinfo {year}
  {2017})}\BibitemShut {NoStop}%
\bibitem [{\citenamefont {Thiele}(1973)}]{thiele1973}%
  \BibitemOpen
  \bibfield  {author} {\bibinfo {author} {\bibfnamefont {A.~A.}\ \bibnamefont
  {Thiele}},\ }\href {\doibase 10.1103/physrevlett.30.230} {\bibfield
  {journal} {\bibinfo  {journal} {Physical Review Letters}\ }\textbf {\bibinfo
  {volume} {30}},\ \bibinfo {pages} {230} (\bibinfo {year} {1973})}\BibitemShut
  {NoStop}%
\bibitem [{\citenamefont {Schütte}\ and\ \citenamefont
  {Garst}(2014)}]{schutte2014magnon}%
  \BibitemOpen
  \bibfield  {author} {\bibinfo {author} {\bibfnamefont {C.}~\bibnamefont
  {Schütte}}\ and\ \bibinfo {author} {\bibfnamefont {M.}~\bibnamefont
  {Garst}},\ }\href {\doibase 10.1103/physrevb.90.094423} {\bibfield  {journal}
  {\bibinfo  {journal} {Physical Review B}\ }\textbf {\bibinfo {volume} {90}},\
  \bibinfo {pages} {094423} (\bibinfo {year} {2014})}\BibitemShut {NoStop}%
\bibitem [{\citenamefont {Lin}\ \emph {et~al.}(2014)\citenamefont {Lin},
  \citenamefont {Batista},\ and\ \citenamefont {Saxena}}]{lin2014internal}%
  \BibitemOpen
  \bibfield  {author} {\bibinfo {author} {\bibfnamefont {S.-Z.}\ \bibnamefont
  {Lin}}, \bibinfo {author} {\bibfnamefont {C.~D.}\ \bibnamefont {Batista}}, \
  and\ \bibinfo {author} {\bibfnamefont {A.}~\bibnamefont {Saxena}},\ }\href
  {\doibase 10.1103/physrevb.89.024415} {\bibfield  {journal} {\bibinfo
  {journal} {Physical Review B}\ }\textbf {\bibinfo {volume} {89}},\ \bibinfo
  {pages} {024415} (\bibinfo {year} {2014})}\BibitemShut {NoStop}%
\bibitem [{\citenamefont {Petrova}\ and\ \citenamefont
  {Tchernyshyov}(2011)}]{petrova2011}%
  \BibitemOpen
  \bibfield  {author} {\bibinfo {author} {\bibfnamefont {O.}~\bibnamefont
  {Petrova}}\ and\ \bibinfo {author} {\bibfnamefont {O.}~\bibnamefont
  {Tchernyshyov}},\ }\href {\doibase 10.1103/physrevb.84.214433} {\bibfield
  {journal} {\bibinfo  {journal} {Physical Review B}\ }\textbf {\bibinfo
  {volume} {84}},\ \bibinfo {pages} {214433} (\bibinfo {year}
  {2011})}\BibitemShut {NoStop}%
\bibitem [{\citenamefont {Mochizuki}(2012)}]{mochizuki2012}%
  \BibitemOpen
  \bibfield  {author} {\bibinfo {author} {\bibfnamefont {M.}~\bibnamefont
  {Mochizuki}},\ }\href {\doibase 10.1103/physrevlett.108.017601} {\bibfield
  {journal} {\bibinfo  {journal} {Physical Review Letters}\ }\textbf {\bibinfo
  {volume} {108}},\ \bibinfo {pages} {017601} (\bibinfo {year}
  {2012})}\BibitemShut {NoStop}%
\bibitem [{\citenamefont {D{\'{\i}}az}\ \emph {et~al.}(2020)\citenamefont
  {D{\'{\i}}az}, \citenamefont {Hirosawa}, \citenamefont {Klinovaja},\ and\
  \citenamefont {Loss}}]{Diaz2020}%
  \BibitemOpen
  \bibfield  {author} {\bibinfo {author} {\bibfnamefont {S.~A.}\ \bibnamefont
  {D{\'{\i}}az}}, \bibinfo {author} {\bibfnamefont {T.}~\bibnamefont
  {Hirosawa}}, \bibinfo {author} {\bibfnamefont {J.}~\bibnamefont {Klinovaja}},
  \ and\ \bibinfo {author} {\bibfnamefont {D.}~\bibnamefont {Loss}},\ }\href
  {\doibase 10.1103/physrevresearch.2.013231} {\bibfield  {journal} {\bibinfo
  {journal} {Physical Review Research}\ }\textbf {\bibinfo {volume} {2}},\
  \bibinfo {pages} {013231} (\bibinfo {year} {2020})}\BibitemShut {NoStop}%
\bibitem [{\citenamefont {Weber}\ \emph {et~al.}(2022)\citenamefont {Weber},
  \citenamefont {Fobes}, \citenamefont {Waizner}, \citenamefont {Steffens},
  \citenamefont {Tucker}, \citenamefont {Böhm}, \citenamefont {Beddrich},
  \citenamefont {Franz}, \citenamefont {Gabold}, \citenamefont {Bewley},
  \citenamefont {Voneshen}, \citenamefont {Skoulatos}, \citenamefont {Georgii},
  \citenamefont {Ehlers}, \citenamefont {Bauer}, \citenamefont {Pfleiderer},
  \citenamefont {Böni}, \citenamefont {Janoschek},\ and\ \citenamefont
  {Garst}}]{Weber_2022}%
  \BibitemOpen
  \bibfield  {author} {\bibinfo {author} {\bibfnamefont {T.}~\bibnamefont
  {Weber}}, \bibinfo {author} {\bibfnamefont {D.~M.}\ \bibnamefont {Fobes}},
  \bibinfo {author} {\bibfnamefont {J.}~\bibnamefont {Waizner}}, \bibinfo
  {author} {\bibfnamefont {P.}~\bibnamefont {Steffens}}, \bibinfo {author}
  {\bibfnamefont {G.~S.}\ \bibnamefont {Tucker}}, \bibinfo {author}
  {\bibfnamefont {M.}~\bibnamefont {Böhm}}, \bibinfo {author} {\bibfnamefont
  {L.}~\bibnamefont {Beddrich}}, \bibinfo {author} {\bibfnamefont
  {C.}~\bibnamefont {Franz}}, \bibinfo {author} {\bibfnamefont
  {H.}~\bibnamefont {Gabold}}, \bibinfo {author} {\bibfnamefont
  {R.}~\bibnamefont {Bewley}}, \bibinfo {author} {\bibfnamefont
  {D.}~\bibnamefont {Voneshen}}, \bibinfo {author} {\bibfnamefont
  {M.}~\bibnamefont {Skoulatos}}, \bibinfo {author} {\bibfnamefont
  {R.}~\bibnamefont {Georgii}}, \bibinfo {author} {\bibfnamefont
  {G.}~\bibnamefont {Ehlers}}, \bibinfo {author} {\bibfnamefont
  {A.}~\bibnamefont {Bauer}}, \bibinfo {author} {\bibfnamefont
  {C.}~\bibnamefont {Pfleiderer}}, \bibinfo {author} {\bibfnamefont
  {P.}~\bibnamefont {Böni}}, \bibinfo {author} {\bibfnamefont
  {M.}~\bibnamefont {Janoschek}}, \ and\ \bibinfo {author} {\bibfnamefont
  {M.}~\bibnamefont {Garst}},\ }\href {\doibase 10.1126/science.abe4441}
  {\bibfield  {journal} {\bibinfo  {journal} {Science}\ }\textbf {\bibinfo
  {volume} {375}},\ \bibinfo {pages} {1025} (\bibinfo {year}
  {2022})}\BibitemShut {NoStop}%
\bibitem [{\citenamefont {Ogawa}\ \emph {et~al.}(2015)\citenamefont {Ogawa},
  \citenamefont {Seki},\ and\ \citenamefont {Tokura}}]{ogawa2015ultrafast}%
  \BibitemOpen
  \bibfield  {author} {\bibinfo {author} {\bibfnamefont {N.}~\bibnamefont
  {Ogawa}}, \bibinfo {author} {\bibfnamefont {S.}~\bibnamefont {Seki}}, \ and\
  \bibinfo {author} {\bibfnamefont {Y.}~\bibnamefont {Tokura}},\ }\href
  {\doibase 10.1038/srep09552} {\bibfield  {journal} {\bibinfo  {journal}
  {Scientific Reports}\ }\textbf {\bibinfo {volume} {5}},\ \bibinfo {pages} {1}
  (\bibinfo {year} {2015})}\BibitemShut {NoStop}%
\bibitem [{\citenamefont {Onose}\ \emph {et~al.}(2012)\citenamefont {Onose},
  \citenamefont {Okamura}, \citenamefont {Seki}, \citenamefont {Ishiwata},\
  and\ \citenamefont {Tokura}}]{onose2012observation}%
  \BibitemOpen
  \bibfield  {author} {\bibinfo {author} {\bibfnamefont {Y.}~\bibnamefont
  {Onose}}, \bibinfo {author} {\bibfnamefont {Y.}~\bibnamefont {Okamura}},
  \bibinfo {author} {\bibfnamefont {S.}~\bibnamefont {Seki}}, \bibinfo {author}
  {\bibfnamefont {S.}~\bibnamefont {Ishiwata}}, \ and\ \bibinfo {author}
  {\bibfnamefont {Y.}~\bibnamefont {Tokura}},\ }\href {\doibase
  10.1103/physrevlett.109.037603} {\bibfield  {journal} {\bibinfo  {journal}
  {Physical Review Letters}\ }\textbf {\bibinfo {volume} {109}},\ \bibinfo
  {pages} {037603} (\bibinfo {year} {2012})}\BibitemShut {NoStop}%
\bibitem [{\citenamefont {Takagi}\ \emph {et~al.}(2021)\citenamefont {Takagi},
  \citenamefont {Garst}, \citenamefont {Sahliger}, \citenamefont {Back},
  \citenamefont {Tokura},\ and\ \citenamefont {Seki}}]{takagi2021}%
  \BibitemOpen
  \bibfield  {author} {\bibinfo {author} {\bibfnamefont {R.}~\bibnamefont
  {Takagi}}, \bibinfo {author} {\bibfnamefont {M.}~\bibnamefont {Garst}},
  \bibinfo {author} {\bibfnamefont {J.}~\bibnamefont {Sahliger}}, \bibinfo
  {author} {\bibfnamefont {C.~H.}\ \bibnamefont {Back}}, \bibinfo {author}
  {\bibfnamefont {Y.}~\bibnamefont {Tokura}}, \ and\ \bibinfo {author}
  {\bibfnamefont {S.}~\bibnamefont {Seki}},\ }\href {\doibase
  10.1103/PhysRevB.104.144410} {\bibfield  {journal} {\bibinfo  {journal}
  {Phys. Rev. B}\ }\textbf {\bibinfo {volume} {104}},\ \bibinfo {pages}
  {144410} (\bibinfo {year} {2021})}\BibitemShut {NoStop}%
\bibitem [{\citenamefont {Aqeel}\ \emph {et~al.}(2021)\citenamefont {Aqeel},
  \citenamefont {Sahliger}, \citenamefont {Taniguchi}, \citenamefont {M\"andl},
  \citenamefont {Mettus}, \citenamefont {Berger}, \citenamefont {Bauer},
  \citenamefont {Garst}, \citenamefont {Pfleiderer},\ and\ \citenamefont
  {Back}}]{Aqeel2021}%
  \BibitemOpen
  \bibfield  {author} {\bibinfo {author} {\bibfnamefont {A.}~\bibnamefont
  {Aqeel}}, \bibinfo {author} {\bibfnamefont {J.}~\bibnamefont {Sahliger}},
  \bibinfo {author} {\bibfnamefont {T.}~\bibnamefont {Taniguchi}}, \bibinfo
  {author} {\bibfnamefont {S.}~\bibnamefont {M\"andl}}, \bibinfo {author}
  {\bibfnamefont {D.}~\bibnamefont {Mettus}}, \bibinfo {author} {\bibfnamefont
  {H.}~\bibnamefont {Berger}}, \bibinfo {author} {\bibfnamefont
  {A.}~\bibnamefont {Bauer}}, \bibinfo {author} {\bibfnamefont
  {M.}~\bibnamefont {Garst}}, \bibinfo {author} {\bibfnamefont
  {C.}~\bibnamefont {Pfleiderer}}, \ and\ \bibinfo {author} {\bibfnamefont
  {C.~H.}\ \bibnamefont {Back}},\ }\href {\doibase
  10.1103/PhysRevLett.126.017202} {\bibfield  {journal} {\bibinfo  {journal}
  {Phys. Rev. Lett.}\ }\textbf {\bibinfo {volume} {126}},\ \bibinfo {pages}
  {017202} (\bibinfo {year} {2021})}\BibitemShut {NoStop}%
\bibitem [{\citenamefont {Utesov}(2021)}]{Utesov2021}%
  \BibitemOpen
  \bibfield  {author} {\bibinfo {author} {\bibfnamefont {O.~I.}\ \bibnamefont
  {Utesov}},\ }\href {\doibase 10.1103/PhysRevB.103.064414} {\bibfield
  {journal} {\bibinfo  {journal} {Phys. Rev. B}\ }\textbf {\bibinfo {volume}
  {103}},\ \bibinfo {pages} {064414} (\bibinfo {year} {2021})}\BibitemShut
  {NoStop}%
\bibitem [{\citenamefont {Utesov}(2022)}]{Utesov2022}%
  \BibitemOpen
  \bibfield  {author} {\bibinfo {author} {\bibfnamefont {O.~I.}\ \bibnamefont
  {Utesov}},\ }\href {\doibase 10.1103/PhysRevB.105.054435} {\bibfield
  {journal} {\bibinfo  {journal} {Phys. Rev. B}\ }\textbf {\bibinfo {volume}
  {105}},\ \bibinfo {pages} {054435} (\bibinfo {year} {2022})}\BibitemShut
  {NoStop}%
\bibitem [{\citenamefont {Timofeev}\ and\ \citenamefont
  {Aristov}(2022)}]{Timofeev_2022}%
  \BibitemOpen
  \bibfield  {author} {\bibinfo {author} {\bibfnamefont {V.~E.}\ \bibnamefont
  {Timofeev}}\ and\ \bibinfo {author} {\bibfnamefont {D.~N.}\ \bibnamefont
  {Aristov}},\ }\href {\doibase 10.1103/physrevb.105.024422} {\bibfield
  {journal} {\bibinfo  {journal} {Physical Review B}\ }\textbf {\bibinfo
  {volume} {105}},\ \bibinfo {pages} {024422} (\bibinfo {year}
  {2022})}\BibitemShut {NoStop}%
\bibitem [{\citenamefont {Timofeev}\ \emph {et~al.}(2019)\citenamefont
  {Timofeev}, \citenamefont {Sorokin},\ and\ \citenamefont
  {Aristov}}]{Timofeev2019}%
  \BibitemOpen
  \bibfield  {author} {\bibinfo {author} {\bibfnamefont {V.~E.}\ \bibnamefont
  {Timofeev}}, \bibinfo {author} {\bibfnamefont {A.~O.}\ \bibnamefont
  {Sorokin}}, \ and\ \bibinfo {author} {\bibfnamefont {D.~N.}\ \bibnamefont
  {Aristov}},\ }\href {\doibase 10.1134/S0021364019030056} {\bibfield
  {journal} {\bibinfo  {journal} {{JETP} Letters}\ }\textbf {\bibinfo {volume}
  {109}},\ \bibinfo {pages} {207} (\bibinfo {year} {2019})}\BibitemShut
  {NoStop}%
\end{thebibliography}%
\end{document}